\documentclass[aps,rmp,twocolumn,american]{revtex4}
\usepackage[T1]{fontenc}
\usepackage[latin1]{inputenc}
\usepackage{amssymb}
\bibpunct{[}{]}{;}{,}{,}

\makeatletter

\usepackage{pslatex,amsmath,amssymb,amsxtra,amsfonts}
\usepackage{citesort}
\usepackage{babel}
\makeatother
\begin{document}

\title{\Huge{Nonlocality, Bell's Ansatz and Probability}}

\author{A. F. Kracklauer}

\homepage{www.nonloco-physics.000freehosting.com}

\begin{abstract}
Quantum Mechanics lacks an intuitive interpretation, which is the
cause of a generally formalistic approach to its use. This in turn
has led to a certain insensitivity to the actual meaning of many words
used in its description and interpretation. Herein, we analyze carefully
the possible mathematical meanings of those terms used in analysis
of EPR's contention, that Quantum Mechanics is incomplete, as well
as Bell's work descendant therefrom. As a result, many inconsistencies
and errors in contemporary discussions of nonlocality, as well as
in Bell's Ansatz with respect to the laws of probability, are identified.
Evading these errors precludes serious conflicts between Quantum Mechanics
and both Special Relativity and Philosophy.
\end{abstract}
\maketitle

\section{A conundrum: nonlocality}

Obscurity in the meaning, or potential meaning, of the Quantum Mechanical
(QM) wave function led \textsc{Einstein, Padolski} and \textsc{Rosen}
to claim that \textsc{Born}'s interpretation of the modulus squared
of a wave function as a probability of presence, implies that QM is
incomplete.\cite{key-1} \textsc{Bohr} disagreed; but his arguments
are not transparent.\cite{key-2} The position argued by \textsc{Bohr}
is developed more transparently, perhaps, by a line of reasoning beginning
with \textsc{von Neumann} and carried forth by \textsc{Bohm} and
finally expounded elegantly by \textsc{John Bell}. They sought to
either prove, or disprove by example, that completing QM could be
accomplished only at the cost of introducing elements into the envisioned
extended theory more disagreeable even than the putative incompleteness
of QM; in the case of \textsc{Bell} specifically, this feature is
\emph{nonlocality}.\cite{key-3} 

In the end, all of \textsc{Bell}'s considerations on this issue, are
based on the following argument and formula. We paraphrase:

Consider the disintegration of a spin-$0$ Boson into two daughters,
comprising a system described by the singlet state. If now a measurement
made of the spin of one daughter, let it be denoted {}``$A$'',
in the $\vec{a}$ direction, i.e., $\sigma _{A}\cdot \vec{a}$, yields
$A(a)=+1$, then a measurement of {}``$B$'', $\sigma _{BA}\cdot \vec{b}$,
must yield $B(b)=-1$. These outcomes are, in the usual formulation
of QM, fundamentally probabilistic, either side for any given pair
could be $\pm 1$, the only constraint here is the subsequent deterministic
anti-correlation of its partner.

Completion of QM, \textsc{Bell} takes it, means that there should
exist further, as yet undiscerned, variables $\lambda $, which, were
they knowable and measurable, would enable the deterministic prediction
of the outcomes, thereby eliminating the statistical character now
intrinsic to QM. Thus, \textsc{Bell} writes altered symbols denoting
the outcomes of each measurement for the proposed experiment: $A(\vec{a},\, \lambda )$
and $B(\vec{b},\, \lambda )$; that is, they are given the form of
deterministic functions of their arguments. More importantly, he takes
it, that by virtue of \emph{locality,} what happens at station {}``$A$''
cannot depend on the value of $b$, the settings of the measurement
device at station {}``$B$''; and of course, visa versa. Then, without
further ado, \textsc{Bell} set forth his now famous Ansatz, that the
expectation of their product should be, \emph{if it is to be local
and realistic,} given by \begin{equation}
P(\vec{a},\, \vec{b})=\int d\lambda \, \rho (\lambda )A(\vec{a},\, \lambda )B(\vec{b},\, \lambda ),\label{eq:1}\end{equation}
 where $\rho (\lambda )$, is a putative distribution of the still
undiscerned%
\footnote{The standard term \emph{``hidden,''} instead of \emph{undiscerned},
presupposes some external agent other than incapability or inattention
of observers; a logically unjustified covert assumption.%
} variables. Finally, \textsc{Bell} notes, that by the algorithms of
QM, this should equal the expectation:\begin{equation}
<\vec{\sigma }_{A}\cdot \vec{a},\, \vec{\sigma }_{B}\cdot \vec{b}>=-\vec{a}\cdot \vec{b}.\label{eq:2}\end{equation}
 On the basis of Eq. (\ref{eq:1}), \textsc{Bell} (and others) derived
several versions of his renowned inequalities which have been shown
empirically to be incompatible with Eq. (\ref{eq:2}). That is, the
data taken in actual experiments is accurately predicted by Eq. (\ref{eq:2}),
but does not satisfy those inequalities derived by \textsc{Bell} et
al. from Eq, (\ref{eq:1}). As satisfying these inequalities is considered
a test for the assumptions going into their derivation, disagreement
means that both QM and experiments do not meet the assumptions. The
final conclusion usually drawn from all this is, that QM is incompatible
with a \emph{local} extended or deeper theory eliminating the statistical
aspects of quantum theory, because locality is thought to have been
the only disputable hypothetical input into the derivation of Bell
inequalities.

All of this is very well known and generally considered utterly beyond
dispute.

There are, however, disturbing features in \textsc{Bell}'s Ansatz
(even ignoring, momentarily, sharply critical, contrarian analysis\cite{key-4}),
that are captured by the observation that Eq. (\ref{eq:1}) exactly
as \textsc{Bell} wrote it, cannot be found in treatises on probability
and statistics! This, we believe, very revealing curiosity, is the
motivation for this study.

\section{Back to basics}

The source of this ``curiosity'' must be in the, partially implicit,
relationship between the imputed physics in \textsc{Bell}'s arguments
and its encoding in symbolics for mathematical manipulation. It is
the purpose here to parse precisely such relationships, in particular
to identify any implicit or covert aspects. 

To begin, it is imperative to understand the structure of \textsc{Bell}'s
argument, i.e., its overarching logic, which was to determine the
possibility of extending QM in such a way that its weird features
would be eliminated. That is to say, the \emph{hypothetical} extended
formalism, the goal, ideally should contain no ill-defined, ambiguous
or contradictory aspects. While perhaps not the only option, the most
obvious first choice for searching afield for this ideal structure
would be the domain of classical physics, for which the only known
defect was, that it did not fully explain all observed phenomena,
not that it, itself, is internally contradictory. In any case, it
can be cogently argued, \textsc{Bell} seemed to have assumed implicitly
that the target structure would be fully compatible with known mathematics.
Let us join him now explicitly in this latter assumption.

What this entails is that all the symbols in Eq. (\ref{eq:1}) must
be given meaning in conformity with the practices as presented in
standard texts on probability and statistics. All but the un-discerned
variables, $\lambda $, are operationally defined and, therefore,
already have fully compatible identities devoid of preternatural properties.
Thus, now only for the variables $\lambda $, must it be so arranged
that they not be ascribed properties unrealizable within classical
mathematical physics. This is fulfilled if a device can be found that
can measure their values. If this consideration is not imposed, then
one of \textsc{Bell'}s primary motivations is frustrated at the start.

Now, it is asserted often that according to QM both $A(\vec{a})$
and $B(\vec{b})$ are random variables, that is, they take on the
values $\pm 1$ randomly. $A(\vec{a},\, \lambda )$ and $B(\vec{b},\, \lambda )$
are, therefore, essentially identical to $A(\vec{a})$ and $B(\vec{b})$,
with the difference, that the symbol $\lambda $ in the set of arguments
is a place-holder for that information, which because it is missing
renders $A(\vec{a})$ and $B(\vec{b})$ random (but appropriately
correlated) variables, i.e., were this information available, they
would be deterministic functions. This brings one, however, straight
up against a source of deep ambiguity in the currently used symbolics. 

It is this: in order for Eq. (\ref{eq:2}) to hold, we must be able
also to write\begin{equation}
<\vec{\sigma }_{A}\cdot \vec{a}>=\int d\lambda \, \rho (\lambda )A(\vec{a},\, \lambda ).\label{eq:3}\end{equation}
 While this seems innocent enough, it begs an issue, namely what should
$A(a,\, \lambda )$ actually signify? Confusion arises immediately
when it is recognized that in experiments no measuring apparatus actually
outputs readings of $\pm 1$ (regardless of whatever units are impugned,
but overlooked); rather, in optical experiments%
\footnote{As a practical matter, it seems, one need consider only optical experiments,
as several obstacles have prevented carrying out convincing experiments
with particles having `entangled spin.'%
}, the photodetectors simply register the excitation of a photoelectron
in a detector behind some kind of filter. The numerical values `$\pm 1$',
then, can be, at most, just labels for the channels in which the photodetectors
``fired,'' indicating that a signal arrived at the detector that passed
through the filter standing in front of it. The mesh size of the filter,
as it were, is indicated by the label $a$ or $b$. In turn, Eq. (\ref{eq:3})
must be understood as an expression of the ratio:
\begin{equation}
<\sigma \cdot a>=\frac{Q_{a}}{N},\label{eq4}
\end{equation}
where $Q_{a}$ then is the \emph{number of times} a photoelectron
appeared in the channel in which the polarizer-filter setting was
given by $a$, in $N$ trials, in the limit $N\rightarrow \infty $.
This is, of course, virtually the very definition of a probability. 

At this point some additional insight can be gained by imagining carrying
out an experiment explicitly. Consider an optical experiment in which
the source is a parametric down conversion crystal (PDC, type II)
producing signals anticorrelated with respect to polarization.\cite{key-5}
In each arm of the setup there is then a polarizer with its axis set
in some direction (e.g., $a$ or $a'$, etc.) perpendicular to the
line of flight. Behind each polarizer there is a photodetector. Typically,
the source intensity is set so low that the likelihood is that there
is only one `photon' in each signal of the pair, that is, operationally
this is understood to mean that only one photoelectron appears in
the detector circuitry for each arm. Now, in the course of an experiment,
what happens exactly? First, for the sake of simplicity, let us assume
that we know somehow that the first pair has been generated, i.e,
$n=1$. Then, a check is made at each photodetector to see if a photoelectron
was registered. If now the left, $A$, polarizer's axis was in the
$\vec{a}_{1}$ direction, and the $B$ polarizer in the $\vec{b_{1}}$
direction, then the data taken for this run would be (where, for example,
a hit was registered at $A$ but not $B$): $n=1$, $\vec{a}=\vec{a_{1}}:\, yes$;
$\vec{b}=\vec{b_{1}}:\, no$. To take this data, three things must
be determined: 1) that a pair was generated, 2) the settings of the
polarizers and 3) the output of the photodetectors.%
\footnote{In the experiment described here a `no-hit' at a detector in one arm
is usually understood as a positive detection in a companion channel,
customarily labeled `$-1$.' By using a polarizing beam splitter,
with each output face feeding a separate detector, both outcomes can
be registered as positive events, thereby reducing ambiguity. Such
refinements do not affect the points being made in the present argument,
however.%
}

Of course, what is not known in this case is the precise polarization
of the signals comprising the pair as emitted at the source but before
they reach the polarizer-filters. The polarizer settings can be known
because they are inputs into measuring devices under the control of
the experimenter who selects their orientation before the pair is
generated at the source. The effect of a polarizer is described by
\textsc{Malus}' Law, and the digitized response of detectors is a
consequence of the nature of individual photoelectron generation at
very low stimulus level.

Seen this way, it is absolutely indisputable that such detector settings
have no effect on the source, and, therefore, have no affect on the
pair of signals before they enter the polarizers. They do, of course,
have an effect on whether particular signals will pass the polarizers
and trigger `hits' in the photodetectors, but they do so independently,
without collaboration. They do have, therefore, a direct contributing
affect on those photo-detection events counted for the experiment. 

By continuing the experiment, additional data can be taken for additional
runs $n=2,\, 3,\, \ldots N$ leading to ratios for which the numerator
is the number of photo-detections when $a=\vec{a_{1}}$, or $\vec{a_{2}}$,
and the denominator is the number of runs, $N$, i.e., $n_{a_{1}}/N$,
and likewise on the other side, $n_{b_{1}}/N$. Reasonably, after
a sufficient number of runs, we expect these ratios to converge to
the probabilities of photo-detections under the circumstance of the
experiment. Thus, we may write them as, e.g., $P(a_{1})$.

This is a standard application of probability theory; \textsc{Bell}'s
program was to ask: what information is needed to so extend the quantum
theory of this experiment that it becomes deterministic (and fault
free)? He supposed that the variables denoted $\lambda $ above, when
known, would specify the conditions of the experiment so that the
outcome with each run could be unambiguously determined in advance,
at least in principle. Such a circumstance can be imagined as follows:
First we shall take it that the source has fixed orientation; for
a source exploiting parametric down conversion, this is achieved if
the known axis of the crystal is fixed in a given direction so that
vertically polarized signals are vertical with respect to both the
crystal and laboratory. This implies that horizontal signals will
be likewise, also with respect to the crystal and laboratory. Then
all the information that is needed to render each experimental run
fully deterministic is the polarization orientation of one of the
signals as it departs the source, either vertical or horizontal, as
the other must be complementary. 

In the usual setup, exactly this information is not available to the
experimenter, it is for him, ``hidden.'' But, presumably \textsc{Zeus}
knows it, and, were he the experimenter, he could discern the value
of the `hidden' variable, $\lambda $, such that for him the outcomes
would be deterministic and predictable in advance. For him, the $P(a_{1},\, \lambda )$
becomes then a Kronecker delta function equal to ``yes'' (which means
that a photoelectron is elevated into the conduction band in a detector)
for that value of $\lambda $ that corresponds to a specific event
comprising a particular pair of anticorrelated signals. Further, \textsc{Zeus}
could take complete data, that is, for each pair he could write down,
in addition to the run, $n$, and the settings of the polarizer filters,
$\vec{a}$ and $\vec{b}$, also the values of $\lambda $, for example,
left signal: $0^{\circ }$; right signal: $90^{\circ }$. Moreover,
he would know the intensity of background fluctuations, and whatever
other signals contribute to the generation of a photoelectron in each
channel, right and left, so that he could predict with certainty whether
a photoelectron is in fact to be generated in each arm after passing
a polarizer with a given orientation. By knowing all this information,
\textsc{Zeus} would then also be in position to sort the final data
stream in terms of the values of $\lambda $ into groups of similar
values; and, in that case (and only in that case) by the precepts
of standard probability theory, the factorization evident in Eq. (\ref{eq:1})
with respect to $\vec{A(a,\, \lambda )}$ and $B(\vec{b,\, \lambda )}$,
\emph{}could be carried out. In the language of a probability theorist,
\textsc{Zeus} can ``screen off'' the variable $\lambda $.

The mortal experimenter, however, with no means of knowing the values
of $\lambda $, cannot sort the data into groups within which the
value of $\lambda $ is constant. Thus, even though there are underlying
specifiable causes, insofar as for the mortal analyst they are in
fact unknowable, sorting on $\lambda $ is \emph{for him} impossible,
therefore the factors $A(a|\lambda )$ and $B(b|\lambda )$ in the
form of data are \emph{for him} sortable only with respect to the
values of $\vec{a}$ and $\vec{b}$. At this stage the purpose of
an experiment is to study the correlations seen in the pattern of
joint hits or detections given the settings, and because the source
was selected on the basis of providing correlated pairs of output
signals, it is absolutely necessary to use Bayes' formula, namely:\begin{equation}
P(a,\, b\, |\lambda )=A(a|b,\, \lambda )B(b|\lambda ),\label{eq:5}\end{equation}
 for the product in the integrand in Eq. (\ref{eq:1}). \cite{key-6}
The outcomes of the measurements are correlated (indeed, observing
this correlation is the purpose of the experiment), because the inputs
were correlated, and this fact necessitates using Eq. (\ref{eq:5})
whatever the significance of $\lambda $. In this application, the
factor $A(a|b,\, \lambda )$ is no longer an independent probability,
but a \emph{conditional} probability, which answers the question:
what is the probability of a hit at station $A$ when set to $\vec{a}$,
given that a hit \emph{was} registered at station $B$ when its polarizer
is set to $\vec{b}$. In this case, carrying along the symbol $\lambda $
is just a reminder that additional but unknowable information could
obviate the need for statistical analysis---as is always true. Again,
there is, contrary to \textsc{Bell}'s argument, no implication whatsoever
that the settings $\vec{a}$ and $\vec{b}$ affect the signals as
generated at the source before they encounter polarizers, or that
they affect each other during detection just because the term $A(a|b,\, \lambda )$
has the parameter $\vec{b}$ within within its complement of arguments.
All this means is that the total polarizer filter setup, before the
signal pair was even generated, was so chosen that it gives the \emph{conditional}
probability of correlated signals for these particular polarizer setting
combination given by $a$ and $b$. If the source pair is not compatible
with the preselected polarizer settings, then as filters, the polarizers
do not pass the signals to the detectors to generate detection events
that can be counted; an inappropriate pulse pair simply does not contribute
to the data stream; it is rejected by the logic of the coincidence
circuitry as spurious background, e.g., as an ``accidental.'' Mathematically,
the distinct form of Eq. (\ref{eq:5}), reflects the restriction for
the association of factors $A(a|\lambda )$ with $B(b|\lambda )$
so as to take into account that not just any outcome at station $A$
can be a cofactor with a particular value of $B(b|\lambda )$ because
of the \emph{correlation} invested in the pair by a \emph{common cause}
within the past light cones of both measuring stations. In EPR-B experiments
this correlation results from limits imposed on the pairs, as generated
at the source, to being \emph{anti-aligned} in terms of polarization,
i.e., to being anticorrelated. \emph{}In any case, the role that $a$
and $b$ have in the symbolics is totally passive; and, association
of these variables with the determination of any property of the signals
generated at the source is, as simply a matter of probability theory,
misplaced. The appearance of a parameter for distinct events or remote
objects in a \emph{conditional} probability does not imply the existence
of a continuing connection (vice structural compatability) of any
kind, much less specifically non-local interaction.\cite{key-6}

This restriction in the pattern of cofactors is expressed in the symbolics
by employing the formula Eq. (\ref{eq:5}). By correctly employing
this formula, one finds as an immediate consequence, that derivations
of Bell inequalities do not proceed. On the other hand, it is just
as clear that they do proceed when, as a restricted case, Eq. (\ref{eq:1})
does pertain, that is, when there is no \emph{correlation} between
the factors $A(a|\lambda )$ and $B(b|\lambda )$. Therefore, on the
basis of these considerations, Bell inequalities are applicable only
to ensembles of \emph{uncorrelated} pairs. Clearly then, testing them
with \emph{correlated} signal pairs must lead to invalid
conclusions.\cite{key-7,key-8,key-9}

\section{Double conundrum: irreality}

This story, so far, is too simple for direct application to QM! There
is an additional and serious complication. It is brought into the
matter, although \textsc{Bell} did not make explicit mention of it
in most of his papers, by \textsc{von Neumann}'s measurement theory
with its `projection hypothesis.'\cite{key-10}

For reasons (that we shall try to analyze below), it is taken often
that the state of a single signal pair in QM is given by:\begin{equation}
<l,\, r>=\frac{1}{\sqrt{2}}(|v,\, h>-|h,\, v>),\label{eq:6}\end{equation}
where $l$ and $r$ indicate the left and right arm of an EPR experiment,
and $v$ and $h$ indicate which polarization the signal sent into
the respective arm is to have. This state, for reasons clear from
the QM analysis of spectroscopy data, is called the `singlet state'
and is, as indicated, to be comprised of two mutually exclusive possibilities,
i.e., it is ``irreal.'' According to the orthodox, `Copenhagen' interpretation
of QM, which is based of the presumption of the completeness of QM,
this is the state \emph{actually} (in the full sense of ontology)
describing the signal pair as they depart the source, but before they
trigger detections. At the detectors, however, what is known as \textsc{``von
Neumann}'s measurement theory'' which includes the ``projection hypothesis''
is invoked to account for the fact that a detection on either side
always finds only one of the possibilities; it is asserted, by authority
of this theory, that measurement itself projects the state onto the
base states such that only one outcome (vertical, say) is realized
in one arm. Further then, by symmetry, the state of the signal in
the other arm is also determined by collapse (e.g., horizontal). This
`projection,' or `collapse' of this wave packet, is considered to
transpire instantaneously regardless of the separation of the measuring
stations on the two sides of the experiment; which is, again, as is
very well known, a violation of the principle of Relativity according
to which \emph{no interaction} can transpire faster than the speed
of light.%
\footnote{Arguments to the effect that this transmission cannot be used to communicate
and therefore that it does not violate Relativity, can be challenged
on the grounds that Relativity constrains all \emph{interaction,}
not just communication. Communication, after all, is just modulation
on interaction of some sort. %
}

For the moment, let us not question any of this. Instead let us see
what consequences imposing these considerations on those in the previous
section might have.

Recall to start, that above we observed that Bayes' formula:\begin{equation}
P(a,\, b)=P(a|b)P(b),\label{eq:7}\end{equation}
does not imply any particlar type of \emph{physical} interaction between
the stations $A$ and $B$. In standard probability theory, it is
never a point of contest, because it is simply taken that the correlation
was invested by a ``common cause'' in the past, so that no violation
of Special Relativity is involved. The fact is, however, this need
not be true to comply with the logic of conditional probabilities.
In other words, from strictly the mathematical point of view, it would
be acceptable to call on a `projection hypothesis' to resolve any
essential ambiguity of singlet-type states; so, mathematically this
issue is no obstacle.

The logic of \textsc{Bell}'s analysis is absolutely opposed, however.
\textsc{Bell}'s intention when conceiving of his ``proof,'' excluded
insinuating, at the meta-level where the inequalities are being derived,
any hypothesis not found in classical, local and realistic physics
as it was understood before the discovery of QM, where the interpretation
issues of QM do not exist. His explicit purpose was to examine the
question of the existence of a covering theory that has just that
structure exploited by classical, pre-quantum theories. His tactic
taken in the proof of what has become known as a ``theorem,'' although
he himself never so designated his argument as such, was to assume
that a \emph{problem free} super theory exists in principle, and then
derive constraints from it that then should percolate down to the
lower quantum theory as it is understood nowadays to see if they are
compatible with what exists, at that coarser level. The point here
is, that this motivation precludes altogether bringing the \textsc{von
Neumann} measurement theory with its `projection hypothesis' (and
all else unique to QM) into the higher level insofar as this structure
is both unknown in classical physics and in violation of Special Relativity
on its face. Of course, one might dispute the necessity of wave function
collapse even at the level of QM, but that is a separate issue.

This admonition deserves strong emphasis. Many authors discuss the
EPR conundrum and \textsc{Bell}'s analysis without scrupulous attention
to and explicit mention of \textsc{Bell}'s overarching logic and thereby
fall into implicit ambiguity. They bring these special and non-classical
features, occasionally even directly at the hypothetical meta-level,
implicitly into the story as if they were somehow germane and fully
legitimate, contrary to \textsc{Bell}'s primary objective.

\section{Underlying algebra}

Given that the two signals in an EPR-B experiment can be in two states
each, there will then be four possible combinations for the pair.
The fact that for these experiments the paired signals must be anticorrelated
eliminates the two even combinations, leaving the two pair-states:
\begin{equation}
|v,\, h>\textrm{and}\: \textrm{|}h,\, v>.\label{eq:8}\end{equation}
In so far as these are mutually exclusive states, they may be regarded
as orthogonal vectors spanning a two-dimensional space. This is, of
course, a formal association, the utility of which depends on full
compatibility of the physical and mathematical structures, a proposition
to be examined. Nevertheless, accepting this formality allows considering
a rotation of the axis of the two-dimensional space to get the superposition
states:\begin{equation}
\frac{1}{\sqrt{2}}(|v,\, h>+\textrm{|}h,\, v>),\: \textrm{and}\: \frac{1}{\sqrt{2}}(|v,\, h>-\textrm{|}h,\, v>).\label{eq:9}\end{equation}
Although a perfectly legitimate vector space operation, as an ontological
statement about the physical objects associated with the vectors,
it is ``otherworldly.'' Consider, for example, a room in a building
in which a coordinate system is defined to be such that one wall is
the abscissa, $x$, and a perpendicular wall the ordinate $y$. Here
each dimension is associated with a material object, a wall; but,
in the rotated system obtained by employing the transformations $x'=x+y;\; y'=x-y$,
there is no material association with $x'$, indeed; it runs through
the middle of the room where there is no wall. The same principles
apply to the Hilbert space comprising the solution space to a Sturm-Liouville
differential equation. Moreover, even before this consideration can
be brought to bear, for a Hilbert space there is no natural, intrinsic
association of the basis vectors for the solution space with ontological
substantive entities. Therefore, \emph{a fortiori}, there is no implicit
association with ontological objects in transformed bases. Such associations
must be individually established and physically justified for each
application, presumably by deliberate matching superpositions of eigen
functions with physically valid boundary or initial conditions; it
cannot  be expected that ontologically valid states fall out of a
formalism mindlessly. 

Superpositions (sums) of mutually exclusive objects, such as Eqs.
(\ref{eq:9}), do not enjoy ontological existence with respect to
ordinary logic. How is it, then, that such ``irreal'' objects are
taken as ontologically acceptable within the QM formalism?

Eqs. (\ref{eq:9}) are deduced usually, i.e., in textbooks, by considering
symmetry requirements implied by the indistinguishability of perfectly
identical particles. The explication virtually always begins with
the observation that, because such particles cannot be individually
distinguished, the Hamiltonian for a system comprising two such identical
particles, for example, must be perfectly symmetric with respect to
the exchange of the coordinates of the particles. (See, e.g., \cite{key-11}.)
Thereafter, however, in spite of general similarities, the hypothetical
inputs are delineated and laid out with enormous variety and often
imprecision. 

The first problem arising here is due to the fact that the application
of this logic to electromagnetic pulses or signals with different
states of polarization ignores the fact that polarization states are
derived from classical electrodynamics; no QM is involved, so that
the Hamiltonian for this structure is classical, and the indistinguishability
of identical entities is not an issue. Classical entities, described
by classical Hamiltonians, always can be identified and distinguished
in principle, so that this consideration should not be brought up
in the first instance.\cite{key-12}

Then, the demonstration that state vectors or wave functions for the
system as a whole need be symmetric or antisymmetric, typically tacitly
assumes that physical (i.e., ontologically valid) states are eigen
vectors of the Hamiltonian. Although this assumption is very widely
made by authors on QM, it is still dubious. Eigen states of the quantized
harmonic oscillator, for example, do not oscillate; an alarmingly
embarrassing feature! The argument that this is a ``mystery'' endemic
to the microword of QM is seriously undermined by \textsc{Moyal}'s
observation that the eigen states of the dynamical equation of a Markoff
process, even in classical physics, do not yield positive definite
Wigner densities.\cite{key-13} Thus, the failure of the eigen states
of the Schr\"odinger equation for the harmonic oscillator to yield
everywhere positive definite Wigner densities cannot be taken as
evidence of ineluctable \emph{quantum} incomprehensibility at a microscopic
level. Insofar as coherent states both oscillate and yield positive
definite Wigner densities, perhaps it should be taken that only they
are legitimately identified as ontologically meaningful states.

Likewise, the eigen functions of the wave equation, the trigonometric
functions, being finite on an infinite domain, are also clearly non
physical; indeed these states are considered unphysical simply because
they cannot be normalized. It is arguably reasonable, therefore, to
presume that physically relevant solutions for the equations of mathematical
physics are restricted to only those combinations of eigen functions
that satisfy some physically realizable initial or boundary conditions. 

This reasoning brings analysis of the physical significance of Eqs.
(\ref{eq:9}) to the question: what do in fact the hypothetical inputs
into the arguments introducing such superpositions actually imply
about their ontological or physical meaning? To begin, let us ignore
the technical problems just mentioned, and reason simply from ``ground
up.'' It is a tautology that either a wave function pertains to a
single system (perhaps comprised of several parts), or it does not.
If it does not, then the superposition can be understood easily. In
this case, a wave function can be taken to specify not the named object,
but rather the preparation procedure that yields this object for observation.\cite{key-14}
The presence, then, of mutually exclusive sub-items or terms, is not
problematic, it means just that the procedure can produce either of
the options at separate times such that each is a separate member
of an ensemble. This option appears to be entirely problem free, but
still is the minority view. Because it implies that QM is incomplete,
it supports the thrust of the \textsc{Einstein, Podolsky} and \textsc{Rosen}
incompleteness argument\textsc{,} which, supposedly, was rebutted
successfully by \textsc{Bohr,} and so nowadays is rejected \textsc{}routinely;
but, we argue, falsely\textsc{.} 

On the other hand, taking a wave function to pertain to single systems
necessitates introducing the projection hypothesis to account for
the fact that no superposition of mutually exclusive observable states
is ever seen in the laboratory (ignoring the logical absurdity of
the expectation that it could). In addition, it seems that the singlet-triplet
state structure is not universally applicable. In particular for EPR
setups, where it is taken that the outcomes pairwise are deterministically
anticorrelated, two of the triplet states are not available at the
start for contribution to the relative frequencies of the observations,
thereby calling into question the use of the term `singlet state'
for these applications, i.e., the `singlet' distinction is germane
only \emph{v\'is-a-v\'is} the presence of `triplet' states which
engender three contrasting signals. (Below, we shall give an even
better reason for this contention.) Moreover, the logic used to motivate
the conception of such states, i.e., the indistinguishability of identical
particles, does not pertain to polarization states, they are always
distinguishable.

Perversely, the essentially probabilistic nature ascribed to wave
functions precludes empirically determining the validity of the their
completeness and the incumbent projection hypothesis. A single data
point (for a single pair) in an EPR experiment resolves nothing. Likewise,
Heisenberg uncertainty cannot be tested with a single data point.
A numerically significant sample of data points is needed to permit
doing the statistical analysis implicit in these very concepts. In
plain talk, there is no measurement to be made on a single system
that can verify the presumption that it is \emph{completely} described
by its quantum wave function. By all logic, all qualities of wave
functions can be fixed only using statistical analysis of an ensemble.
\emph{}Here it would seem, therefore, that \textsc{Occam's} razor
should leap into action to preempt \textsc{Popper}'s objection to
introducing untestable hypothesis into scientific theories, in this
case, being the completeness of wave functions with the then incumbent
`projection hypothesis.'

\section{Underlying physics}

Eqs. (\ref{eq:9}), whilst being superpositions of mutually exclusive
outcomes, were not taken into the tool kit of QM just because they
enjoy algebraic legitimacy. They have very useful applications, apparently
the first of which was considered by \textsc{Heisenberg} in 1926,
where he applied the then new techniques introduced by QM to explain
the spectrum for multi-electron atoms.\cite{key-15} Helium, with
two electrons, is the simplest case; and, its first excited state---with
either one or the other of its two electrons in an excited $2P$-state,
while its partner is in the $1S$-ground state---is degenerate, because
there are two possible combinations leading to the same circumstance.

Using degenerate perturbation theory one finds directly but tediously,
that the eigen vectors for the coupled case have the form of Eqs.
(\ref{eq:9})., and that the correction to the energy i.e, $E_{\textrm{correction}}=E_{\textrm{Coulomb}}\pm E_{\textrm{exchange}}$,
where the first term is given by:\begin{equation}
E_{\textrm{Coulomb}}=\int u_{1}^{*}(r_{1})u_{1}(r_{1})\frac{e^{2}}{r_{1}-r_{2}}u_{2}^{*}(r_{2})u_{2}(r)_{2},\label{eq:11}\end{equation}
 and can be interpreted classically as the expectation of the Coulomb
interaction of the charge distributions for both electrons. The second
term is the result of the so-called ``exchange force:''\begin{equation}
E_{\textrm{exchange}}=\int u_{1}^{*}(r_{1})u_{2}(r_{2})\frac{e^{2}}{r_{1}-r_{2}}u_{2}^{*}(r_{2})u_{1}(r_{1}),\label{eq:12}\end{equation}
 where in this case the interaction is for electrons in an atom and
is given by $V=e^{2}/(r_{1}-r_{2})$. Clearly, as an indisputable
historical matter, this is the logic and structure that led to the
introduction of such states into QM in the first instance. 

An absolutely crucial condition for application of degenerate perturbation
theory is that there exists a physical interaction between the electrons.
When interaction is absent, then both the term for Coulomb interaction
and that for the exchange force vanish, as they do for noninteracting
light pulses (doubly true with different polarizations). Then the
eigen values need no corrections and revert to simple products of
the unperturbed eigen values, and the eigen functions are then simple
(factorable) products of the unperturbed eigen functions. It is the
interaction between the electrons that induces correlations spoiling
the factorability. Seen in these terms, it is the interaction that
spoils the statistical independence of the wave functions for the
electrons. In other words, the interaction induces the correlations
between the electrons that mandate introduction of conditional probabilities
and therefore \textsc{Bayes}' formula, Eq. (\ref{eq:7}). 

The form of Eqs. (\ref{eq:9}) can be taken as the structurally simplest
form that captures the nonfactorability that is inevitable by cause
of interaction. Once again, however, there is nothing in degenerate
perturbation analysis that requires that eigen functions be valid
ontological states. They can be just the most elementary, nonfactorable
form; ontological states must be those superpositions of such states
that satisfy initial or boundary conditions. 

When there is no interaction, then degenerate perturbation theory
does not lead to states of this form. Thus, it would seem that EPR/Bell
analysis cannot call on analysis of two polarization states because
electromagnetic signals of differing polarization, according to standard
electrodynamics, do not interact. Because they do not interact, both
the `Coulomb' and exchange term are zero, and the logic leading to
the `singlet' states, as a superposition of mutually exclusive options,
is not applicable. Perhaps one can go a step further and question
even whether the so far hypothetical experiments on particles with
spin would be appropriate, as in EPR experiments, the particles comprising
a pair are separated by relatively immense macroscopic distances,
and the technicalities of the generation of such states at the source,
where interaction does take place, is immaterial for purposes of testing
EPR's hypothesis. Of course, the logic for the anticorrelated polarizations
remains intact; it is just the logic for the singlet state \emph{as
a complete ontological entity} that is set aside. 

Moreover, it is often pointed out that the superposition states correspond
to the mechanical analogue involving two coupled oscillators for which
there are two `eigen modes,' i.e., oscillation both in and out of
phase. When coupled, the oscillators can exchange energy and slosh
back and forth between these modes. This `classical' model, by example,
indirectly supports the understanding of wave functions as being incomplete.

Once again, the identification of eigenstates with ontological states
is a matter for specific examination. Obviously, when the two states
(for electrons or whatever) are coupled, they influence each other
so that the individual states are no longer statistically independent.
The state functions for the system then cannot be the simple product
of the state functions for the parts, and are so rendered nonfactorable.
Eqs. (\ref{eq:9}) can be taken as the simplest form satisfying this
requirement; but ontological states, then, would be superpositions
of such states that satisfy the relevant physically specified boundary
or initial conditions.

Finally, we note that spectral analysis of absorptions and emissions
from atoms cannot be taken as evidence that the ontological states
of the electrons between such absorbtions or emissions are given by
the eigen functions. That these spectra comprise multiple lines, leaves
open the interpretation that the absorption or emission itself was
comprised of a superposition, and only appears to be distinct lines
because the total signal is being `spectrally analyzed' by the optical
instruments used for observation.

\section{Rotational invariance of the singlet state}

The state represented by Eq. (\ref{eq:6}), i.e., the singlet state:\begin{equation}
|l,\, r>=\frac{1}{\sqrt{2}}(|v,\, h>-|h,\, v>),\label{eq:6a}\end{equation}
 is said to be rotationally invariant, which is meant to say that
if the individual states $|v>_{l,r}$ and $|h>_{l,r}$ are expressed
in terms of axes rotated about the fixed wave vector, $\vec{k}$,
e.g., $|v>_{l}=x_{1}\cos \theta +y_{1}\sin \theta $, etc., then the
system state or wave function preserves its form, namely:\begin{equation}
|1,\, 2>=\frac{1}{\sqrt{2}}(|x_{1},\, y_{2}>-|y_{1},\, x_{2}>).\label{eq:10}\end{equation}

This equivalence is, nevertheless, still ambiguous. It could mean
that statistically all averages and moments calculated with both expressions
are equal, which would mean that as a statistical expression, it is
invariant. Beyond this, however, a physical interpretation can be
imposed on the equivalence to the effect that not only are the averages
equal, but actually the individual separate states in both resolutions
are the same, at least insofar as the individual anticorrelation is
deterministic in both resolutions. This extra physical assertion is
an additional hypothetical input that is independent of the mathematics
as necessitated by the statistics. It must be independently verified
by observation. Symbolic manipulations transforming from Eq. (\ref{eq:6a})
to Eq. (\ref{eq:10}) do not depend on or address this matter.

Rotational invariance of singlet-type states as a \emph{physical} assumption
appears to be virtually universally accepted, although sometimes implicitly, in
physics literature. Indeed, it is an essential ingredient in at least one
illustration of the mysteries of QM, \cite{key-16}, and indeed, if true, leads
to mathematical inconsistencies rendering QM indeed mysterious. It appears to
be functionally equivalent, at least in spirit, to the assumption that QM is
complete; i.e., that a wave function pertains to individual systems, in EPR
experiments: to individual signal pairs. For the signals used in EPR-B
experiments employing polarization ``entanglement,'' empirical evidence
collected by this writer, albeit not at the single `photon' level, contradicts
individual, as opposed to statistical, deterministic anticorrelation, thereby
supporting only statistical rotational invariance. \footnote{Of course, such
`multiple photon' observations are necessary but not sufficient support for the
contention that wave functions are not complete.}  The separate states of
polarization are deterministically anticorrelated only in the basis for which
the axis of the PDC crystal is parallel or perpendicular to the axis of
polarizer. On the other hand, it has been verified by simulation that
statistical rotational invariance is fully valid in the sense that arbitrary
rotations introduced into signals for EPR-B type experiments do not affect the
statistical analysis or final determination of correlation functions.
\cite{key-4}

On occasion, rotational invariance is taken even to mean ``spherical
invariance,'' which means that transformations to an arbitrary direction in
space, not just rotations about the wave vector, leaving the form of Eq.
(\ref{eq:6a}) invariant. This mathematical fact is taken to imply yet another
physical hypothetical input, namely, that spin is quantized in all directions
at once, not just in the direction of the magnetic field (which is required to
reveal the existence of spin at all). Obviously, with respect to spin, this is
an absolutely untestable proposition, as it is impossible to have magnetic
fields in more than one direction at a point simultaneously, making this a pure
metaphysical proposition, even oxymoronic. With respect to polarization, the
implied physics suffers the same difficulty. Indeed a wave vector has physical
significance in only one direction; coordinate transformations from the `alias'
point of view are not the issue. \cite{key-12}

\section{Entanglement verses correlation}

According to the Born interpretation, QM state functions give the
probability of presence as their modulus squared. Since the operation
of `squaring' does not affect factorability, it seems there should
be a direct relationship between nonfactorability of wave functions
(a.~k.~a.: entanglement) and that of the probability densities derived
therefrom (native correlation). Nevertheless, they are widely held
to be fundamentally distinct. The natural question is: how?

One answer appears to be: entangled states are those that violate
Bell inequalities. It is said that, in this case, there are `quantum'
correlations `stronger' than admitted by any classical definition.
This definition is a derived property rather than a primitive
characteristic. Moreover, 
it presupposes the validity of the derivation of Bell Inequalities,
a proposition that can be accepted only by overlooking issues delineated
above.

In fact a better grounded reason is that a distinction between these
concepts exists because of those hypothetical elements ultimately
necessitating the `projection hypothesis.' These elements, in turn,
as discussed above, are necessitated by the presumption that QM is
complete, that the wave function for an object is its deepest ontological
manifestation, and that no finer information than what is given by
a wave function is possible, even though, as argued above, this leads
to consideration of `irreal' states (a.k.a ``cat states'').

The internal consistency of the `projection hypothesis' may be testable.
This may have been achieved already using biprisms with electron beams.
In these devices an electron beam is sliced by a negatively charged
wire arranged perpendicular to its propagation direction and passing
through the middle. Upon passing the wire, each half of the beam is
repelled somewhat, so that the two half beams diverge slightly. Then,
downstream, a second parallel wire, but charged positively, draws
the two diverging partial beams together so that they meet on a registration
screen, where interference of their de Broglie waves is observed.\cite{key-17}
Now, each of these wires plays the role of an optical instrument,
but still they are contrivances created by the experimenter which
both act on, and react to, the passing beam particles. In principle,
these wires are `measuring' devices in that some beam properties could
be read out by observing fluctuations induced in the current in the
wire as caused by the passage of the electron beam. As such, these
observations should then, according to the precepts of QM (that is,
with \textsc{von Neumann}'s contribution), precipitate wave collapse.
This in turn should prevent subsequent wave-like behavior of the beam.
Interference seen on the screen shows that collapse did not occur,
the beam exhibited wave-like behavior downstream from these `measurements.'

A separate recent experiment, originally proposed by \textsc{Karl
Popper,} proves the same point for the wave function for systems comprised
of two correlated particles that eventually become widely separated,
as in EPR experiments.\cite{key-18} In this experiment, one beam
of particles (usually photons, but it could be electrons) is sent
through a slit, and the diffraction observed. The correlated
beam on the other side is, in contrast, not sent through a slit, but
observed for diffraction with an identical set of detectors. If wave
function collapse, as envisioned to occur in EPR experiments, happens,
then, although there is no slit on the latter side, it still should
exhibit the same spacial pattern of detections, because its wave function
should be ``collapsed (diffracted) in sympathy with its partner beam.
Observation shows that nothing of this sort happens.\cite{key-19}
The beam that does not pass through a slit, carries on as if nothing
occurred, ignoring the evolution of its partner. 

The conclusion from such observations must be: wave function collapse
does not occur. In turn, entanglement beyond conventional correlation
is an imaginary artifact. Whatever correlations exist among members
of a system described by a wave function, they are identical to those
among classical objects; the projection hypothesis is groundless.

\section{Continuous variable versions of EPR tests}

One of the difficulties mentioned above in experiments to test Bell
inequalities is based on the fact that \textsc{Bohm}'s change of venue from
phase space to qubit space (polarization or spin) introduces unintended, covert
hypothesis.\cite{key-20} The essence of this problem is that while phase space
\emph{can be} quantized, and thereby made non-commuting, polarization (qubit)
space cannot be modified, i.e.,  ``quantized;'' it is already non commuting by
virtue of its geometric structure. Qubit spaces are locked down; for them there
is no option of having either classical (taken to be commuting) or quantum (non
commuting) structure. 

This objection might be evaded by suitable experiments in phase space.
It appears that straight forward `EPR' formulations, are not practical,
however; clever designs are essential. In recent times such proposals
have been made, usually described in terms of ``continuous variables,''
although, as such, \emph{continuity} by itself is not really important;
a `quantum' venue is. On the other hand, continuity complicates the
issue as the usual forms of Bell Inequalities are inadequate for non
discrete variables; a new discriminator is required. \cite{key-21}

The tactic taken most often to find a discrimination criterion
is based on the observation that the correlation
of the ordinary and extraordinary output signals from a PDC crystal
implies that for the transversal components, the EPR stipulations
hold, that is: $x_{o}+x_{e}=0$ and $k_{o}-k_{e}=0$. Thus, appropriate
measurements of diffraction effects on such signals can be used to
probe these features.

The logic of the discriminators proposed in the literature%
\footnote{Excepting discrimination criteria based directly on the validity of
Bell inequalities, e.g., \cite{key-22}.%
} is based on the following assertion: whereas the variances of the
outputs of a PDC are constrained by Heisenberg Uncertainty:\begin{equation}
(\varDelta (x_{o}+x_{e}))^{2}(\varDelta (p_{o}-p_{e}))^{2}\geq 1/4,\label{eq:13}\end{equation}
in fact, QM allows perfect correlations:\begin{equation}
(\varDelta (x_{o}+x_{e}))^{2}(\varDelta (p_{o}-p_{e}))^{2}\geq 0.\label{eq:14}\end{equation}

Experiments then consist of measuring the dispersions seen in these
signals to show that Eq. (\ref{eq:13}) is not satisfied, which is
taken to mean that an ineluctable quantum phenomenon, entanglement
vice `classical correlation,' has been observed.

In point of fact, however, Eq. (\ref{eq:14}) is \emph{not} dictated
by principles unique to QM, but by conservation principles which are
just as valid classically. Moreover, Eq. (\ref{eq:13}), when considered
strictly in a classical venue, cannot be an expression of Heisenberg
Uncertainty---a quantum principle---but just statistical dispersion,
which in principle can be reduced indefinitely. If applied to electromagnetic
signals, Eq. (\ref{eq:13}), even at the `single photon level,' should
be no more that the classical bandwidth limit. 

Most significantly, this claim makes no reference to ``locality,''
the crux of the matter for \textsc{Bell}'s considerations. 

On this basis, it appears that these arguments turn the usual logic
on its head.

\section{Conclusions}

The points made above offer several explanations for the observation
noted in the introduction, that \textsc{Bell}'s Ansatz, Eq. (\ref{eq:1}),
cannot be found in treatises on statistics and probability. To begin,
there is misleading notation; \textsc{Bell} used a comma to separate
the independent arguments, whereas `hidden' variables, by definition
would be conditioning parameters, and, as such, in the notation customary
in works on probability, are separated from independent variables
by a vertical bar. This malapropos turn of the pen appears to have
been an important facilitating element in the general misconstrual
of \textsc{Bell}'s analysis. Once this defect is corrected, it is
a short leap to the understanding of the necessity for applying \textsc{Bayes}'
formula; a leap apparently made first by \textsc{Jaynes. \cite{key-7}}

And, once it is clear that Bell Inequalities cannot be derived using
\textsc{Bayes'} formula, the issue of nonlocality is rendered moot. This, in
turn, resolves one conflict between two fundamental theories of modern
physics---a conflict that on the face of it has the character typical of
small, technical misunderstandings. This is only reinforced by the observation
that there is \emph{no} empirical evidence for nonlocality; that which has
been taken as such, is in fact just an interpretation imposed indirectly on
statistics derived from non kinematic data, but as argued herein, incorrectly.
The main conflict between QM and General Relativity remains, however. The
energy density of the quantized ground state of the free electromagnetic
field, i.e., the ``quantum vacuum,'' is at least $120$ orders of magnitude
larger than allowed by cosmological constant considerations. 

A similarly perplexing philosophical issue brought to the discussion
of the nature of the interpretation of QM is that concerning the ontological
status of states, such as the singlet state, ostensibly comprised
of the superposition of mutually exclusive entities. This particular
issue, seemingly, has caused only mild discomfort among physical scientists,
apparently since it is in conflict `only' with philosophical considerations,
not major physics theories. Nevertheless, it is auspicious, if only
symbolically, that properly understood fundamental physics theories
do not encompass gross conflict with the foundations of the enlightenment
and the scientific revolution. Rejecting the `completeness hypothesis' achieves
just that.

Finally, based on the analysis presented above, it is arguable that the reasoning
behind `measurement theory' and the `projection hypothesis' is fully
disputable, and dispensable; and, that there are coherent alternatives
qualifying for \textsc{Occam}'s approval. 

\bibliography{unsrt}

\end{document}